
%
%
\def\unlockat{\catcode`\@=11}
\def\lockat{\catcode`\@=12}
\unlockat
\def\d@f@ult{} \newif\ifamsfonts \newif\ifafour
\def\m@ssage{\immediate\write16}  \m@ssage{}
\m@ssage{hep-th preprint macros.  Last modified 16/10/92 (jmf).}
\message{These macros work with AMS Fonts 2.1 (available via ftp from}
\message{e-math.ams.com).  If you have them simply hit "return"; if}
\message{you don't, type "n" now: }
\endlinechar=-1  
\read-1 to\@nswer
\endlinechar=13
\ifx\@nswer\d@f@ult\amsfontstrue
    \m@ssage{(Will load AMS fonts.)}
\else\amsfontsfalse\m@ssage{(Won't load AMS fonts.)}\fi
\message{The default papersize is A4.  If you use US 8.5" x 11"}
\message{type an "a" now, else just hit "return": }
\endlinechar=-1  
\read-1 to\@nswer
\endlinechar=13
\ifx\@nswer\d@f@ult\afourtrue
    \m@ssage{(Using A4 paper.)}
\else\afourfalse\m@ssage{(Using US 8.5" x 11".)}\fi
\nonstopmode
%
%

\font\twelverm=cmr12
\font\ninerm=cmr9
\font\sixrm=cmr6
\font\fourteenbf=cmbx12 scaled\magstep1
\font\twelvebf=cmbx12
\font\ninebf=cmbx9
\font\sixbf=cmbx6
\font\fourteeni=cmmi12 scaled\magstep1      \skewchar\fourteeni='177
\font\twelvei=cmmi12                        \skewchar\twelvei='177
\font\ninei=cmmi9                           \skewchar\ninei='177
\font\sixi=cmmi6                            \skewchar\sixi='177
\font\fourteensy=cmsy10 scaled\magstep2     \skewchar\fourteensy='60
\font\twelvesy=cmsy10 scaled\magstep1       \skewchar\twelvesy='60
\font\ninesy=cmsy9                          \skewchar\ninesy='60
\font\sixsy=cmsy6                           \skewchar\sixsy='60
\font\fourteenex=cmex10 scaled\magstep2
\font\twelveex=cmex10 scaled\magstep1

\ifamsfonts
   \font\ninex=cmex9
   
   \font\sixex=cmex7 at 6pt
   
\else
   \font\ninex=cmex10 at 9pt
   
   \font\sixex=cmex10 at 6pt
   
\fi
\font\fourteensl=cmsl10 scaled\magstep2
\font\twelvesl=cmsl10 scaled\magstep1

\font\sevensl=cmsl10 at 7pt
\font\sixsl=cmsl10 at 6pt

\font\fourteenit=cmti12 scaled\magstep1
\font\twelveit=cmti12

\font\fourteentt=cmtt12 scaled\magstep1
\font\twelvett=cmtt12
\font\fourteencp=cmcsc10 scaled\magstep2
\font\twelvecp=cmcsc10 scaled\magstep1

\ifamsfonts
   
\else
   
\fi
\newfam\cpfam
\font\fourteenss=cmss12 scaled\magstep1
\font\twelvess=cmss12
\font\tenss=cmss10
\font\niness=cmss9

\font\sevenss=cmss8 at 7pt
\font\sixss=cmss8 at 6pt
\newfam\ssfam
\newfam\msafam \newfam\msbfam \newfam\eufam
\ifamsfonts
 \font\fourteenmsa=msam10 scaled\magstep2
 \font\twelvemsa=msam10 scaled\magstep1
 \font\tenmsa=msam10
 \font\ninemsa=msam9
 \font\sevenmsa=msam7
 \font\sixmsa=msam6
 \font\fourteenmsb=msbm10 scaled\magstep2
 \font\twelvemsb=msbm10 scaled\magstep1
 \font\tenmsb=msbm10
 \font\ninemsb=msbm9
 \font\sevenmsb=msbm7
 \font\sixmsb=msbm6
 \font\fourteeneu=eufm10 scaled\magstep2
 \font\twelveeu=eufm10 scaled\magstep1
 \font\teneu=eufm10
 \font\nineeu=eufm9
 
 \font\seveneu=eufm7
 \font\sixeu=eufm6
 \def\hexnumber@#1{\ifnum#1<10 \number#1\else
  \ifnum#1=10 A\else\ifnum#1=11 B\else\ifnum#1=12 C\else
  \ifnum#1=13 D\else\ifnum#1=14 E\else\ifnum#1=15 F\fi\fi\fi\fi\fi\fi\fi}
 \def\hexmsa{\hexnumber@\msafam}
 \def\hexmsb{\hexnumber@\msbfam} 
\fi
\newdimen\b@gheight             \b@gheight=12pt
\newcount\f@ntkey               \f@ntkey=0
\def\f@m{\afterassignment\samef@nt\f@ntkey=}
\def\samef@nt{\fam=\f@ntkey \the\textfont\f@ntkey\relax}
\def\rm{\f@m0 }
\def\mit{\f@m1 }
\def\cal{\f@m2 }
\def\it{\f@m\itfam}
\def\sl{\f@m\slfam}
\def\bf{\f@m\bffam}
\def\tt{\f@m\ttfam}
\def\caps{\f@m\cpfam}
\def\ssf{\f@m\ssfam}
\ifamsfonts
 \def\msa{\f@m\msafam}
 \def\msb{\f@m\msbfam} 
 \def\eu{\f@m\eufam}
\else
  \let\eu=\bf
\fi
\def\fourteenpoint{\relax
    \textfont0=\fourteencp          \scriptfont0=\tenrm
      \scriptscriptfont0=\sevenrm
    \textfont1=\fourteeni           \scriptfont1=\teni
      \scriptscriptfont1=\seveni
    \textfont2=\fourteensy          \scriptfont2=\tensy
      \scriptscriptfont2=\sevensy
    \textfont3=\fourteenex          \scriptfont3=\twelveex
      \scriptscriptfont3=\tenex
    \textfont\itfam=\fourteenit     \scriptfont\itfam=\tenit
    \textfont\slfam=\fourteensl     \scriptfont\slfam=\tensl
      \scriptscriptfont\slfam=\sevensl
    \textfont\bffam=\fourteenbf     \scriptfont\bffam=\tenbf
      \scriptscriptfont\bffam=\sevenbf
    \textfont\ttfam=\fourteentt
    \textfont\cpfam=\fourteencp
    \textfont\ssfam=\fourteenss     \scriptfont\ssfam=\tenss
      \scriptscriptfont\ssfam=\sevenss
    \ifamsfonts
       \textfont\msafam=\fourteenmsa     \scriptfont\msafam=\tenmsa
         \scriptscriptfont\msafam=\sevenmsa
       \textfont\msbfam=\fourteenmsb     \scriptfont\msbfam=\tenmsb
         \scriptscriptfont\msbfam=\sevenmsb
       \textfont\eufam=\fourteeneu     \scriptfont\eufam=\teneu
         \scriptscriptfont\eufam=\seveneu \fi
    \samef@nt
    \b@gheight=14pt
    \setbox\strutbox=\hbox{\vrule height 0.85\b@gheight
                                depth 0.35\b@gheight width\z@ }}
\def\twelvepoint{\relax
    \textfont0=\twelverm          \scriptfont0=\ninerm
      \scriptscriptfont0=\sixrm
    \textfont1=\twelvei           \scriptfont1=\ninei
      \scriptscriptfont1=\sixi
    \textfont2=\twelvesy           \scriptfont2=\ninesy
      \scriptscriptfont2=\sixsy
    \textfont3=\twelveex          \scriptfont3=\ninex
      \scriptscriptfont3=\sixex
    \textfont\itfam=\twelveit    
    \textfont\slfam=\twelvesl    
      \scriptscriptfont\slfam=\sixsl
    \textfont\bffam=\twelvebf     \scriptfont\bffam=\ninebf
      \scriptscriptfont\bffam=\sixbf
    \textfont\ttfam=\twelvett
    \textfont\cpfam=\twelvecp
    \textfont\ssfam=\twelvess     \scriptfont\ssfam=\niness
      \scriptscriptfont\ssfam=\sixss
    \ifamsfonts
       \textfont\msafam=\twelvemsa     \scriptfont\msafam=\ninemsa
         \scriptscriptfont\msafam=\sixmsa
       \textfont\msbfam=\twelvemsb     \scriptfont\msbfam=\ninemsb
         \scriptscriptfont\msbfam=\sixmsb
       \textfont\eufam=\twelveeu     \scriptfont\eufam=\nineeu
         \scriptscriptfont\eufam=\sixeu \fi
    \samef@nt
    \b@gheight=12pt
    \setbox\strutbox=\hbox{\vrule height 0.85\b@gheight
                                depth 0.35\b@gheight width\z@ }}
\twelvepoint
%
%
\baselineskip = 15pt plus 0.2pt minus 0.1pt 
\lineskip = 1.5pt plus 0.1pt minus 0.1pt
\lineskiplimit = 1.5pt
\parskip = 6pt plus 2pt minus 1pt
\interlinepenalty=50
\interfootnotelinepenalty=5000
\predisplaypenalty=9000
\postdisplaypenalty=500
\hfuzz=1pt
\vfuzz=0.2pt
\dimen\footins=24 truecm 
\ifafour
 \hsize=16cm \vsize=22cm
\else
 \hsize=6.5in \vsize=9in
\fi
%
%
\skip\footins=\medskipamount
\newcount\fnotenumber
\def\clearfnotenumber{\fnotenumber=0} \clearfnotenumber
\def\fnote{\global\advance\fnotenumber by1 \generatefootsymbol
 \footnote{$^{\footsymbol}$}}
\def\fd@f#1 {\xdef\footsymbol{\mathchar"#1 }}
\def\generatefootsymbol{\iffrontpage\ifcase\fnotenumber
\or \fd@f 279 \or \fd@f 27A \or \fd@f 278 \or \fd@f 27B
\else  \fd@f 13F \fi
\else\xdef\footsymbol{\the\fnotenumber}\fi}
%
%
\newcount\secnumber \newcount\appnumber
\def\clearappnumber{\appnumber=64} \def\clearsecnumber{\secnumber=0}
\clearsecnumber \clearappnumber
\newif\ifs@c 
\newif\ifs@cd 
\s@cdtrue 
\def\unsectioned{\s@cdfalse\let\section=\subsection}
\newskip\sectionskip         \sectionskip=\medskipamount
\newskip\headskip            \headskip=8pt plus 3pt minus 3pt
\newdimen\sectionminspace    \sectionminspace=10pc
\def\Titlestyle#1{\par\begingroup \interlinepenalty=9999
     \leftskip=0.02\hsize plus 0.23\hsize minus 0.02\hsize
     \rightskip=\leftskip \parfillskip=0pt
     \advance\baselineskip by 0.5\baselineskip
     \hyphenpenalty=9000 \exhyphenpenalty=9000
     \tolerance=9999 \pretolerance=9000
     \spaceskip=0.333em \xspaceskip=0.5em
     \fourteenpoint
  \noindent #1\par\endgroup }
\def\titlestyle#1{\par\begingroup \interlinepenalty=9999
     \leftskip=0.02\hsize plus 0.23\hsize minus 0.02\hsize
     \rightskip=\leftskip \parfillskip=0pt
     \hyphenpenalty=9000 \exhyphenpenalty=9000
     \tolerance=9999 \pretolerance=9000
     \spaceskip=0.333em \xspaceskip=0.5em
     \fourteenpoint
   \noindent #1\par\endgroup }
\def\spacecheck#1{\dimen@=\pagegoal\advance\dimen@ by -\pagetotal
   \ifdim\dimen@<#1 \ifdim\dimen@>0pt \vfil\break \fi\fi}
\def\section#1{\cleareqnumber \s@ctrue \global\advance\secnumber by1
   \par \ifnum\the\lastpenalty=30000\else
   \penalty-200\vskip\sectionskip \spacecheck\sectionminspace\fi
   \noindent {\caps\enspace\S\the\secnumber\quad #1}\par
   \nobreak\vskip\headskip \penalty 30000 }
\def\undertext#1{\vtop{\hbox{#1}\kern 1pt \hrule}}
\def\subsection#1{\par
   \ifnum\the\lastpenalty=30000\else \penalty-100\smallskip
   \spacecheck\sectionminspace\fi
   \noindent\undertext{#1}\enspace \vadjust{\penalty5000}}

\def\appendix#1{\cleareqnumber \s@cfalse \global\advance\appnumber by1
   \par \ifnum\the\lastpenalty=30000\else
   \penalty-200\vskip\sectionskip \spacecheck\sectionminspace\fi
   \noindent {\caps\enspace Appendix \char\the\appnumber\quad #1}\par
   \nobreak\vskip\headskip \penalty 30000 }
\def\ack{\par\penalty-100\medskip \spacecheck\sectionminspace
   \line{\fourteencp\hfil ACKNOWLEDGEMENTS\hfil}%
\nobreak\vskip\headskip }
\def\refs{\begingroup \par\penalty-100\medskip \spacecheck\sectionminspace
   \line{\fourteencp\hfil REFERENCES\hfil}%
\nobreak\vskip\headskip \frenchspacing }
\def\endrefs{\par\endgroup}
%
%
\newif\iffrontpage \frontpagefalse
\headline={\hfil}
\footline={\iffrontpage\hfil\else \hss\twelverm
-- \folio\ --\hss \fi }
%
%
\newskip\frontpageskip \frontpageskip=12pt plus .5fil minus 2pt
\def\titlepage{\global\frontpagetrue\hrule height\z@ \relax
               \pubblock\relax }
\def\endtitlepage{\vfil\break\clearfnotenumber\frontpagefalse}
\def\title#1{\vskip\frontpageskip\Titlestyle{\caps #1}\vskip3\headskip}
\def\author#1{\vskip.5\frontpageskip\titlestyle{\caps #1}\nobreak}
\def\and{\par\kern 5pt \centerline{\sl and}}

\def\authors{\vskip\frontpageskip\noindent}
\def\address#1{\par\kern 5pt\titlestyle{\it #1}}
\def\andaddress{\par\kern 5pt \centerline{\sl and} \address}
\def\addresses{\vskip\frontpageskip\noindent\interlinepenalty=9999}
\def\abstract#1{\par\dimen@=\prevdepth \hrule height\z@ \prevdepth=\dimen@
   \vskip\frontpageskip\spacecheck\sectionminspace
   \centerline{\fourteencp ABSTRACT}\vskip\headskip
   {\noindent #1}}

\def\email#1{\fnote{\tentt e-mail: #1\hfill}}

%
%

%

%

%

%
%
\newcount\refnumber \def\clearrefnumber{\refnumber=0}  \clearrefnumber
\newwrite\R@fs                              
\immediate\openout\R@fs=\jobname.refs 
\def\closerefs{\immediate\closeout\R@fs} 
\def\refsout{\closerefs\refs
\unlockat
\input\jobname.refs
\lockat
\endrefs}
\def\refitem#1{\item{{\bf #1}}}
\def\ifundefined#1{\expandafter\ifx\csname#1\endcsname\relax}
\def\[#1]{\ifundefined{#1R@FNO}%
\global\advance\refnumber by1%
\expandafter\xdef\csname#1R@FNO\endcsname{[\the\refnumber]}%
\immediate\write\R@fs{\noexpand\refitem{\csname#1R@FNO\endcsname}%
\noexpand\csname#1R@F\endcsname}\fi{\bf \csname#1R@FNO\endcsname}}
\def\refdef[#1]#2{\expandafter\gdef\csname#1R@F\endcsname{{#2}}}
%
%
\newcount\eqnumber \def\cleareqnumber{\eqnumber=0}
\newif\ifal@gn \al@gnfalse  
\def\veqnalign#1{\al@gntrue \vbox{\eqalignno{#1}} \al@gnfalse}
\def\eqnalign#1{\al@gntrue \eqalignno{#1} \al@gnfalse}
\def\(#1){\relax%
\ifundefined{#1@Q}
 \global\advance\eqnumber by1
 \ifs@cd
  \ifs@c
   \expandafter\xdef\csname#1@Q\endcsname{{%
\noexpand\rm(\the\secnumber .\the\eqnumber)}}
  \else
   \expandafter\xdef\csname#1@Q\endcsname{{%
\noexpand\rm(\char\the\appnumber .\the\eqnumber)}}
  \fi
 \else
  \expandafter\xdef\csname#1@Q\endcsname{{\noexpand\rm(\the\eqnumber)}}
 \fi
 \ifal@gn
    & \csname#1@Q\endcsname
 \else
    \eqno \csname#1@Q\endcsname
 \fi
\else%
\csname#1@Q\endcsname\fi\global\let\@Q=\relax}
%
%
\newif\ifm@thstyle \m@thstylefalse
\def\mathstyle{\m@thstyletrue}
\def\proclaim#1#2\par{\smallbreak\begingroup
\advance\baselineskip by -0.25\baselineskip%
\advance\belowdisplayskip by -0.35\belowdisplayskip%
\advance\abovedisplayskip by -0.35\abovedisplayskip%
    \noindent{\caps#1.\enspace}{#2}\par\endgroup%
\smallbreak}
\def\m@kem@th<#1>#2#3{%
\ifm@thstyle \global\advance\eqnumber by1
 \ifs@cd
  \ifs@c
   \expandafter\xdef\csname#1\endcsname{{%
\noexpand #2\ \the\secnumber .\the\eqnumber}}
  \else
   \expandafter\xdef\csname#1\endcsname{{%
\noexpand #2\ \char\the\appnumber .\the\eqnumber}}
  \fi
 \else
  \expandafter\xdef\csname#1\endcsname{{\noexpand #2\ \the\eqnumber}}
 \fi
 \proclaim{\csname#1\endcsname}{#3}
\else
 \proclaim{#2}{#3}
\fi}
\def\Thm<#1>#2{\m@kem@th<#1M@TH>{Theorem}{\sl#2}}
\def\Prop<#1>#2{\m@kem@th<#1M@TH>{Proposition}{\sl#2}}
\def\Def<#1>#2{\m@kem@th<#1M@TH>{Definition}{\rm#2}}
\def\Lem<#1>#2{\m@kem@th<#1M@TH>{Lemma}{\sl#2}}
\def\Cor<#1>#2{\m@kem@th<#1M@TH>{Corollary}{\sl#2}}
\def\Conj<#1>#2{\m@kem@th<#1M@TH>{Conjecture}{\sl#2}}
\def\Rmk<#1>#2{\m@kem@th<#1M@TH>{Remark}{\rm#2}}
\def\Exm<#1>#2{\m@kem@th<#1M@TH>{Example}{\rm#2}}
\def\Qry<#1>#2{\m@kem@th<#1M@TH>{Query}{\it#2}}
%
%

%
\def\<#1>{\csname#1M@TH\endcsname}
%
%
\def\ref#1{{\bf [#1]}}
\def\ie{{\it i.e.\/}}
%
%

\def\lapprox{\hbox{\lower3pt\hbox{$\buildrel<\over\sim$}}}
\def\gapprox{\hbox{\lower3pt\hbox{$\buildrel<\over\sim$}}}
\def\quotient#1#2{#1/\lower0pt\hbox{${#2}$}}
\def\fr#1/#2{\mathord{\hbox{${#1}\over{#2}$}}}
\ifamsfonts
 \mathchardef\empty="0\hexmsb3F 
 \mathchardef\lsemidir="2\hexmsb6E 
 \mathchardef\rsemidir="2\hexmsb6F 
\else
 \let\empty=\emptyset
 \def\lsemidir{\mathbin{\hbox{\hskip2pt\vrule height 5.7pt depth -.3pt
    width .25pt\hskip-2pt$\times$}}}
 \def\rsemidir{\mathbin{\hbox{$\times$\hskip-2pt\vrule height 5.7pt
    depth -.3pt width .25pt\hskip2pt}}}
\fi
%
%

%
%
%
%
\def\underrightarrow#1{\vtop{\ialign{##\crcr
      $\hfil\displaystyle{#1}\hfil$\crcr
      \noalign{\kern-\p@\nointerlineskip}
      \rightarrowfill\crcr}}} 
\def\underleftarrow#1{\vtop{\ialign{##\crcr
      $\hfil\displaystyle{#1}\hfil$\crcr
      \noalign{\kern-\p@\nointerlineskip}
      \leftarrowfill\crcr}}}  

\def\comm#1#2{\left[#1\, ,\,#2\right]}
%
\def\der#1#2{{{d #1}\over {d #2}}}
%
%

%
\lockat
%
%

\def\Y{{\ssf Y}}

\let\pb=\anticomm

\def\fr#1/#2{\mathord{\hbox{${#1}\over{#2}$}}}

\def\H{\mathord{\cal H}}
\def\ket|#1>{\mathord{\vert{#1}\rangle}}

\def\ope#1#2{{{#2}\over{\ifnum#1=1 {z-w} \else {(z-w)^{#1}}\fi}}}

\def\corr<#1>{\mathord{\langle #1 \rangle}}
\def\dlb#1#2{\lbrack\!\lbrack#1,#2\rbrack\!\rbrack}

\mathstyle
\overfullrule=0pt
\unsectioned
%

\def\dlb#1#2{\lbrack\!\lbrack#1,#2\rbrack\!\rbrack}

\def\ope#1#2{{{#2}\over{\ifnum#1=1 {x-y} \else {(x-y)^{#1}}\fi}}}

\def\O{{\cal O}}

\def\M{{\ssf M}}

\def\H{{\ssf H}}
\def\G{{\cal G}}
%
\refdef[Eduardo]{E. Ramos, unpublished.}
\refdef[strings]{
B Zwiebach, {\sl Closed string field theory : An introduction.}
MIT-CTP-2206, {\tt hep-th/9305026}.}
\refdef[Kirillov]{
A. A. Kirillov, {\sl Elements of the theory of representations.}
Berlin, Heidelberg, New York: Springer 1975.}

\def\pubblock{ \line{\hfil\twelverm Preprint-QMW-PH-93-17}
               \line{\hfil\twelverm May 1993}
               \line{\hfil hep-th/9305140}}
\titlepage
\title{A symplectic structure for the space of
quantum field theories}
\authors
\hfil{\caps Eduardo Ramos
\email{ramos@v2.ph.qmw.ac.uk\hfil} {\it and}$\,\,$
Oleg A. Soloviev\email{soloviev@v2.ph.qmw.ac.uk\hfil}}
\addresses
{\it Department of Physics, Queen Mary and Westfield College,
Mile End Road, London E1 4NS, UK.}\hfil\break\noindent

\abstract{We use the formal Lie algebraic structure in the ``space''
of hamiltonians provided by equal time commutators to
define a Kirillov-Konstant symplectic structure in the coadjoint
orbits of the associated formal group. The dual is
defined via the natural pairing between operators and states
in a Hilbert space.}

\endtitlepage

\subsection{Introduction}

There has been an increasing interest in the construction of
action functionals in what can be called loosely the ``space of quantum
field theories''. This interest has mainly sprung from the idea that
the configuration space for string field theory is the space of all
two dimensional conformal field theories (\[strings] and references
therein). The fact that
this space, by itself, does not seem to be connected seems to necessitate
working in a bigger space which, itself, contains all conformal field
theories. In this state of affairs, it seems natural to consider as a
candidate for the configuration space the space of all two dimensional
quantum field theories.

But certainly the interest for this subject
is not limited to the string field theoretical approach.
One may conjecture that the existence of hamiltonian
structures in the space of quantum field theories could lead
to a possible hamiltonian approach to  particular kinds of
renormalization group flows. In fact, part of our interest in the
subject came from the observation that the renormalization group
flows for the Sine-Gordon theory are hamiltonian up to two loops
\[Eduardo]. It is reasonable to suppose that if this property
is not just an artifact of low-order perturbation theory the
partition function itself should provide us with a natural hamiltonian
for the flows. The question that follows naturally is how to determine
the nonperturbative expression for the symplectic structure.

Here it will be shown, how
to equip the space of all quantum field theories
(not necessarily two dimensional) with a symplectic structure by
a (hopefully) judicious use of the formal Lie algebraic structure
provided by the equal time commutator algebra.
This infinite-dimensional Lie algebra has a natural pairing induced
by the pairing among operators and states. The coadjoint
action defined by requiring the invariance of this pairing allows
us to define a Kirillov-Konstant two form \[Kirillov] in the
coadjoint orbits of the associated formal group.

Before starting we whish to point out that our construction
is quite formal - in the physicist's sense of the word. Although
we are able to define symplectic structures for
$D+1$ dimensional quantum field theories for arbitrary $D$, we expect
that a rigorous approach to the subject (if at all possible) would be
restricted to $D=0$, {\ie} to ordinary quantum mechanics. Nevertheless,
by defining the quantum field theory  both on a
space lattice and inside a finite box, in order to control ultraviolet and
infrared divergences, the problem becomes a quantum mechanical one.
For which all our manipulations have some hope to be properly defined.
It can then be expected
that a smooth continuum limit can be achieved. It may also
be possible that, via some standard renormalization
procedure, the symplectic structure could be computed perturbatively
in the neighborhood of a free field theory.
Here we adopt the viewpoint that formal manipulations though
sometimes misleading can also be rather
illuminating.

\subsection{The General Setup}

First of all, we should be a little more explicit about what we mean
by the space of all quantum field theories in $D+1$ dimensions. It
will be convenient to start with classical field theories and move
later to the quantum case.

In the classical case we need two bits of information to define our
classical field theory: one kinematical, consisting of
a symplectic phase space which we will denote by $\Y$,
and  the other dynamical, a hamiltonian function on $\Y$. When we will
talk about the space of all classic field theories, we will refer
to the ``space of all hamiltonian functions on $\Y$''. That is,
we will restrict ourselves to a given symplectic structure.
Is this much of a restriction? In the classical
case certainly not. In order to understand this a little better, let
us first fix our attention to the $D=0$ case, {\ie} standard
classical mechanics with a finite number of degrees of freedom. And
let us consider the case in which the configuration space
enjoys only one degree of freedom, say, the position of the particle.
The phase space is two dimensional and by the Darboux theorem it admits
(at least locally) coordinates in which the symplectic structure
becomes the canonical one, {\ie} $\pb{p}{q}=1$. Now the space of all
$0+1$ classical field theories with one degree of freedom will become,
in our terminology, the space of all possible hamiltonian functions in
that phase space. This essentially exhausts the possible dynamical
systems with one degree of freedom. Of course, if we would consider
sytems with infinite number of degrees of freedom some subtleties would
arise. But in the context of our formal approach  these complications
are ignored.

The first step in our
construction should be to define a symplectic phase space which will
be the natural arena for defining the space of field theories, or
equivalently, the space of hamiltonians. Of course, one
can take the approach of taking the starting phase space so big as to
accomodate any possible number of fields or degrees of freedom.
Perhaps, a more ``workable'' approach is to
start with the desired degrees of freedom and consider all
possible associated hamiltonians which fullfil certain symmetry
or regularity properties.

What about the quantum case?  The analogue of the
phase space is a Hilbert space ${\cal H}$, and that of Poisson
brackets is an equal time commutator algebra. Of course in the quantum
case we do not have an analogue of the Darboux theorem, making the choice of
the starting commutator algebra more arbitrary. From now on,
 whe we talk about the space of all quantum field theories of
dimension $D+1$ the reader should keep in mind that we have already
fixed the field content, and the Hilbert space structure.
To fix ideas, one could think
that we have chosen the Hilbert space associated with
a set of free fields with
cannonical commutation relations. Then the space of all quantum
field theories becomes equivalent to the space of all possible
self-adjoint hamiltonians on those free fields,
{\ie} $\H_0 +\H_{pert} $ for all possible perturbations.

In what follows we will completely ignore questions such as how to
properly define the hamiltonian ``operators'', what is their
domain, etc.

\subsection{The Formal-ism}

Let us denote by $\M^D$ the space of all quantum field theories
in $D+1$ dimensions. As explained before we will parametrize
the points of $\M^D$ by their associated Hamiltonian operator $\H$.
A set of local coordinates on $\M^D$ can be defined as follows.
Let us assume, as is usual in formal treatments,
that there is a countable basis for the algebra (or a big enough
subalgebra) of local operators.
If we denote by  $\O_i(x)$, with $i$ belonging to the set of positive
integers, the basis elements for the subalgebra of self-adjoint
operators, any point in $\M^D$ can be written as
$$\H_U\equiv\sum_j \int dx^D\, u^j\O_j(x),\()$$
and the coupling constants $u^j$ provide us with
a set of convenient coordinates.

The space $\M^D$ can be given a natural formal Lie algebraic structure
$$\comm{\H_U}{\H_V}=i\,\H_{\dlb{U}{V}},\(liealg)$$
with
$$\dlb{U}{V}^l=f^l_{jk}u^jv^k,\(liealgdos)$$
where the structure constant coeficients are defined by
$$\comm{\int\O_j}{\O_k(x)}= i\, f^l_{jk}\O_l(x),\()$$
and the commutator is defined through the composition of ``operators''.
The extra factor of $i$ is there to preserve ``self-adjointness''.

The next step is to identify the dual, {\ie} the set of linear
functionals on the formal Lie algebra defined above. The dual
can be identified with $\M^D$
by considering the natural pairing among operators and
states in a Hilbert space. Let us define
$$U(V)=\langle\Omega_U\mid\H_V\mid\Omega_U\rangle,\(pairing)$$
where $\mid\Omega_U\rangle$ is the ground state associated to
the hamiltonian $\H_U$. Notice that, as defined,
the map that assigns to $U$ an element of the dual $U(\cdot )$
is nonlinear,and therefore $U(V)\neq V(U)$ and
$(U+W)(V)\neq U(V) + W(V)$.
Nevertheless, as we will show
now, this coadjoint action turns out to be the natural one.

The coadjoint action is defined by requiring invariance of
\(pairing). For infinitesimal transformations this requirement
reads
$$(U-ad^*_W(U))(V)= U(V) + U(ad_W(V)),\(coadrep)$$
where $ad_W(V)$ is defined to be $\dlb{W}{V}$. It is now simple
to check that if
$$ad^*_W (U) =\dlb{W}{U}\()$$
the pairing is invariant.
In order to show this, let us first compute the right hand side of
\(coadrep)
$$\eqalign{
&U(V) + U(ad_W(V))= U(V)-
i\langle\Omega_U\mid\comm{\H_W}{\H_V}\mid\Omega_U\rangle\cr
=&U(V)+i\sum_p\left(\langle\Omega_U\mid\H_V\mid p\rangle
\langle p\mid\H_W\mid\Omega_U\rangle -
\langle\Omega_U\mid\H_W\mid p\rangle
\langle p\mid\H_V\mid\Omega_U\rangle\right),\cr}\(choruno)$$
where $\sum_p\mid p\rangle\langle p\mid$ stands for a resolution of
the identity in eigenstates of $\H_U$.

The left hand side of
\(coadrep) can now be computed by using first order perturbation
theory. Notice that at first order in $W$
$$\eqalign{
\mid\Omega_{U+\dlb{W}{U}}\rangle =&\mid\Omega_U\rangle
-i\sum_p{{\langle p\mid\comm{\H_W}{\H_U}\mid\Omega_U\rangle}\over
{E_p - E_0}}\mid p\rangle\cr
=&\mid\Omega_U\rangle +i\sum_p\langle p\mid\H_W\mid\Omega_U\rangle
\,\mid p\rangle,\cr}\(chordos)$$
where we have assumed that the ground state is nondegenerate. It is
now a simple algebraic computation to check that
$$\langle\Omega_{U+\dlb{W}{U}}
\mid \H_V\mid\Omega_{U+\dlb{W}{U}}\rangle\()$$
reproduces \(choruno) up to second order terms in $W$.

\subsection{The Coadjoint Orbit Method}

Thanks to our previous construction, it is now easy to define a
$G$-invariant symplectic structure in the coadjoint orbits of
the formal group $G$ associated to the formal Lie algebraic
structure defined by \(liealgdos).

Let us briefly recall the general construction in a way that
will suit our needs. Let $G$ be a Lie
group and $\G$ its Lie algebra. Let us denote by $\G^*$ the space
of linear functionals on $\G$. For the time being we will denote
the elements of $\G$ with greek characters, and the ones of $\G^*$
by latin characters. The coadjoint representation is defined by
requiring invariance of the pairing among elements of $\G^*$ and
$\G$, {\ie} $\forall g\in G$
$$(Ad^*_{g^{-1}}b)(\xi)=b(Ad_g\xi).\(groupad)$$
If we now fix an element $b$ of $\G^*$, we will denote by $O_b$ its
orbit under $G$.

Vector fields on $O_b$ are naturally parametrized by elements of $\G$. Let us
define $\partial_{\xi}\in TO_b$ at a point $a\in O_b$ by
$$\partial_{\xi} f(a) =\der{\ }{\epsilon} f(a +\epsilon\, ad^*_{\xi}(a))
\Bigm|_{\epsilon =0}\, ,\()$$
with $f$ an arbitrary function on $O_b$.
Notice that these vectors are defined up to elements in the stability
subalgebra of $a$, elements  $\eta\in\G$ such that $ad_{\eta}^*(a)=0$.
Therefore $\partial_{\xi +\eta}$ and $\partial_{\xi}$ define the
same tangent vector at that point.

We can now define a symplectic form on any point $a\in O_b$ by
$$\omega (\partial_{\xi},\partial_{\chi})= a (\comm{\xi}{\chi}),
\(coadform)$$
which is obviously antisymmetric and $G$-invariant. It is also
well defined because if $\eta$ belongs to the stability subalgebra
of $a$
$$\eqalign{
\omega (\partial_{\xi +\eta},\partial_{\chi})=&
\omega (\partial_{\xi},\partial_{\chi}) + a(\comm{\eta}{\chi})\cr
=&\omega (\partial_{\xi},\partial_{\chi}) + a( ad_{\eta}(\chi ))\cr
=&\omega (\partial_{\xi},\partial_{\chi}) + (a -ad_{\eta}^*(a))(\chi )
- a(\chi )\cr
=&\omega (\partial_{\xi},\partial_{\chi}).\cr}\()$$

In order to show that $\omega$ is nondegenerate, let us suppose
that there exists a vector $\partial_{\beta}\in TO_b$ such that,
at point $a\in O_b$,
$$\omega (\partial_{\beta},\partial_{\chi})=0\ \ \forall\partial_{\chi}.\()$$
This clearly implies that $\beta$ belongs to the stability
subalgebra\fnote{Notice that in order to show this property we have to
use the linear structure in the dual.}
at that point and therefore parametrizes the zero vector.

Finally, closedness of $\omega$ is easily proved as follows
$$\eqalign{
d\omega (\partial_{\alpha},\partial_{\beta},\partial_{\gamma})=&
\partial_{\alpha}\omega (\partial_{\beta},\partial_{\gamma}) -
\omega (\comm{\partial_{\alpha}}{\partial_{\beta}},\partial_{\gamma})
+ {\rm cyclic\  permutations}\cr
=& \partial_{\alpha} a(\comm{\beta}{\gamma})
- a(\comm{\comm{\alpha}{\beta}}{\gamma})
+ {\rm cyclic\  permutations}\, .\cr}\(deomega)$$
The first term in \(deomega) is zero because of the invariance of
the pairing, while the second term is zero when we sum up over
cyclic permutations because of Jacobi identities in $\G$.

We can now return to our case and define a symplectic form in the
coadjoint orbits of the formal group $G$
by defining $\omega$ at the point $U$ to be
$$\omega (\partial_V,\partial_W)= U(\dlb{V}{W})=
-i\langle\Omega_U\mid\comm{\H_V}{\H_W}\mid\Omega_U\rangle.\()$$

This expression enjoys all the properties described above except
nondegeneracy. This can be seen by considering the one form at the point
$U$ defined by $\omega (\partial_V,\,\cdot\, )$
which is identically zero as long as $\mid\Omega_U\rangle$ is
an eigenstate of $\H_V$, even if $V$ does not belong to the stability
subalgebra of $U$. Notice that the hamiltonians with such a property
form a closed subalgebra.
The problem of characterizing the symplectic leaves of $\omega$
is left open.

\subsection{Final Comments}

We hope to have convinced the reader that the method of coadjoint
orbits can play an important role in the search for a symplectic
structure in the space of all quantum field theories, and that
further investigation of the subject is warranted. Of course,
it is of primary importance to find some simple examples where
this formalism could be applied and where its relevance could be checked.

\ack

We would like to thank Francisco Figueirido, Jose M. Figueroa O'Farrill,
Takashi Kimura, Javier Mas, and Jens Petersen for useful
conversations on the subject.
\bigskip

\refsout

\bye